# Difference-Frequency Chirped-Pulse Dual-comb Generation in the THz Region: Temporal Magnification of the Quantum Dynamics of Water Vapor Lines by >60,000

JASPER R. STROUD[1] AND DAVID F. PLUSQUELLIC[2]

*Applied Physics Division, Physical Measurement Laboratory, National Institute of Standards and Technology, Boulder, CO 80305.*
[1]*jasper.stroud@nist.gov*
[2]*david.plusquellic@nist.gov*

**Abstract**
A new difference-frequency method based on electro-optic phase modulators (EOMs) and two free-running lasers is reported to perform THz dual-comb spectroscopy. A variation of a near-IR interleaving scheme we recently reported has been developed to interleave the EOMs' orders and sidebands, and to map THz comb teeth into the radio-frequency region below 1 MHz. The down-converted comb teeth are shown to have transform limited widths of 1 Hz over a 1 sec time scale. The dual chirp scheme is used to measure the complex line shapes of two water vapor lines below 600 GHz and to temporally magnify the effects of rapid passage by more than 60,000. The method is applicable to any region where difference or sum frequency waves can be generated.

Dual optical frequency combs (DOFC) have served as a powerful tool in numerous application areas by allowing the unique down-conversion and mapping of the comb teeth in the optical domain to the radiofrequency (RF) region for high-throughput detection [1,2,3,4,5]. While typically limited in bandwidth coverage, electro-optic DOFCs (EO-DOFC) have been shown to provide a facile route to modify the comb resolution and bandwidth [6,7,8,9,10,11,12] to optimize the signal-to-noise ratio for a particular application. In recent work, we have demonstrated a new capability of EO-DOFCs generated using a free-running near-IR laser [9,10] to temporally magnify the molecular dynamics of $CO_2$ using dual chirped-pulse waveforms. By applying a differential chirp rate between the signal (SIG) and local oscillator (LO) legs of the interferometer, a magnified view of the rapid passage complex signal response (magnitude and phase) of $CO_2$ was observed in a 2 m cell at pressures up to 13.33 kPa (100 Torr). The transformation of the normal transmission and dispersion line shape profiles to their corresponding rapid passage forms was nearly complete at a magnification ratio of ≈ 150 for scan rates up to ≈ 50 MHz/ns. Key to observing this transformation was the development of interleaving scheme to separate the different orders and sidebands of the EOMs that define the dual combs. In this work, we have adapted this interleaving method for a two-laser multiheterodyne photomixing [13] system to perform dual-comb measurements [14,15,16] on water vapor lines in the THz region.

One element of focus in this work is the further enhancement of the temporal magnification of the coherent quantum dynamics which can be trivially changed by simply reprogramming the chirped waveforms to cover different ranges and durations. Details on the waveform relations can be found in experimental methods. Simple in form, the magnification, $\gamma$, is determined from the ratio of the optical, $\Delta f_{Opt}$, to radio, $\Delta f_{RF}$, frequency comb bandwidths given by,

$$\gamma = \Delta f_{Opt}/\Delta f_{RF} \tag{1}$$

The optimum RF bandwidth used for multi-heterodyne detection often depends on the available source power. For broadband THz photomixers [13,17,18,19], the optical-to-THz conversion efficiency is very low (< 0.01% [13]) with peak output powers typically less than 10 μW. Furthermore, for room-temperature photomixer receivers, the low THz-to-electrical conversion efficiency requires the use of high-gain trans-impedance amplifiers (>$10^5$ V/A) for detection. The need for high gain limits the RF bandwidth to less than a few MHz [5] in contrast to near-IR detector bandwidths which can operate at GHz speeds.

The down-conversion of GHz bandwidths in the THz region to small RF bandwidths is only possible by having narrow comb teeth (<5 Hz) and adequate RF comb resolution (10 Hz - 100 Hz) for the unique mapping to apply. However, once these conditions are met, the temporal magnification factors in Eq. 1 can reach 10's of thousands as we demonstrate in this work. The EO-DOFC technology makes use of off-the-shelf components and provides a robust method for investigations of coherent quantum dynamics without the need for short pulse mode-locked lasers and associated phase locking electronics.

A second element of interest is the development of a new chirped-pulse dual-comb configuration based on two free-running lasers [20,21]. The method takes advantage of the mutual phase coherence at the difference frequencies (DF) generated at the transmitter and receiver to down-convert a THz comb to the RF region [14,15,16]. DF generation gives access to spectral regions defined by the mixed output of the two laser sources and not the electronic bandwidth of the microwave (MW) sources driving the EOMs. As such, this method is expected to apply equally well to techniques for sum frequency generation.

In this letter, the quantum dynamics of water vapor are investigated over a range of pressures to follow the spectral evolution from the normal complex line shape to the rapid passage signal response. For a 10 ms scan, the temporal magnification factor realized in this work is

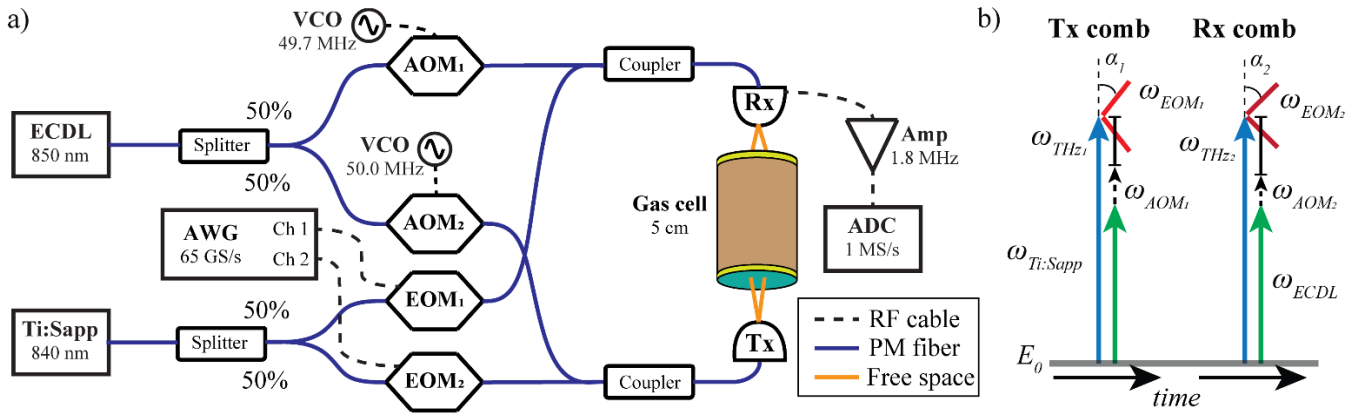

**Figure 1**. a) Schematic of the interleaved DF-EO-DOFC spectrometer. The two lasers used to generate difference frequencies in the THz region include an ECDL and a CW Ti:Sapp laser. Each laser is split equally. The two legs of the ECDL are coupled to AOMs driven near 50 MHz at a difference frequency, $\Delta f_{AOM}$, of 300 kHz. The two legs of the Ti:Sapp laser are coupled to EOMs driven by chirped pulse waveforms generated by the AWG. Light from one AOM and one EOM are combined, and fiber coupled to each of the Rx and Tx low-temperature-grown GaAs photomixers. The THz output beam ($\varphi_D \approx$ 2 mm - 3 mm) from the Tx (biased at 11 V DC) is free space coupled through the sample and focused on the Rx receiver. The electrical signal output from the Rx is amplified in a transimpedance amplifier ($10^6$ V/A) and digitized at 1 MS/s. b) A schematic illustration of the difference frequency system. The ECD (green), is split and shifted with different AOM frequencies (dashed lines) while the Ti:Sapp (blue) is split and sent to two EOMs that are driven by chirped-pulse waveforms (red). THz combs are generated on Tx for transmission and Rx for down-conversion at the difference frequencies of the two lasers. The chirp rates, $\alpha_1$ and $\alpha_2$, are slightly different for the unique mapping to the RF region.

> 60,000 which is more than 400-fold larger than that demonstrated in the previous near-IR study [10]. Two different rotational lines of water are investigated: the $1_{1,0} \leftarrow 1_{0,1}$ ($J_{Ka,Kc}$ notation) transition of $H_2O$ at 556.936 GHz (18.577 cm$^{-1}$) and the $2_{1,1} \leftarrow 2_{0,2}$ transition of $D_2O$ at 403.561 GHz (13.461 cm$^{-1}$) [22]. The observed responses are compared to predictions from the Maxwell-Bloch Equations (MBE) [23,24]. For the $H_2O$ line only, the rapid passage signal response has revealed some intriguing phase flipping quantum dynamics that are shown to depend on pressure.

The difference-frequency electro-optic dual-optical-frequency-comb (DF-EO-DOFC) system is shown in Fig. 1a. The two continuous wave (CW) lasers consist of an external cavity diode laser (ECDL, New Focus, Model 6316) [25] that is amplified to 300 mW using a tapered amplifier (New Focus, TA-7600) and a Ti:Sapphire (Ti:Sapp) ring laser (Coherent 699) pumped with up to 10 W by a solid state Nd:YVO$_4$ laser at 532 nm (Verdi V18, Coherent, Inc.) generating up to 200 mW. The approximate short-term linewidths of the ECDL and Ti:Sapp lasers are near 1 MHz. Uncoated optical wedges are used to pick off small portions of each for wavemeter readout and for long-term drift control (< 1 MHz) using a HeNe laser stabilized reference cavity [26,27]. The scheme for THz wave generation is illustrated in Fig. 1b. The ECDL and Ti:Sapp laser frequencies are represented as vertical green and blue lines, respectively. The added frequency offsets from the acoustic-optic modulators (AOMs) are shown as dashed black lines. The different chirp rates, $\alpha_1$ and $\alpha_2$, and ranges covered by the EOM's chirped pulses are shown for the ($\pm$) sidebands as tilted red lines. The THz frequency ranges generated on the photomixer receiver (Rx) and transmitter (Tx) are indicated with solid black lines. The optical and THz waves are mixed on the Rx to generate the down-converted RF comb (not shown). Although other configurations are possible, the one shown maintains equal powers and fiber lengths to preserve the mutual phase coherence between the two lasers.

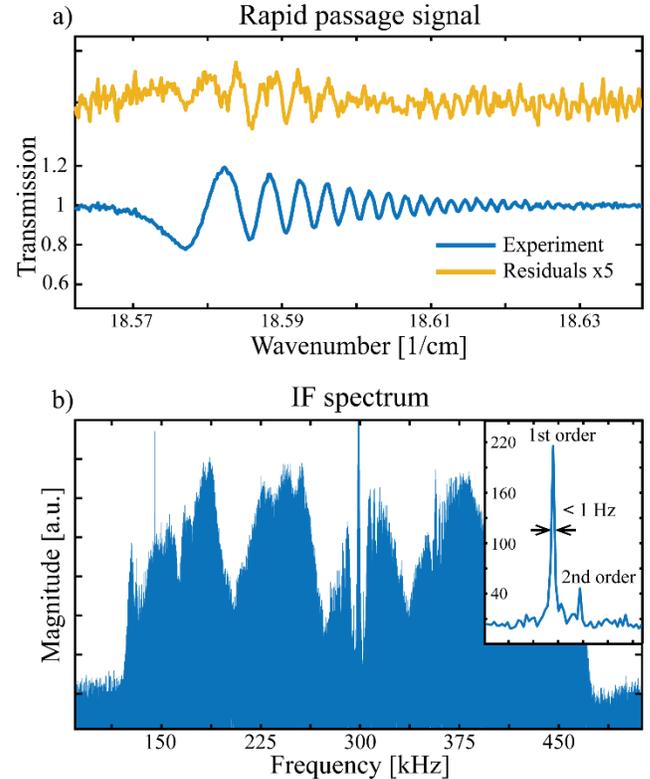

**Figure 2**. Experimental results showing the magnitude of the rapid passage signal response of the $H_2O$ transition at 557 GHz ($1_{10} \leftarrow 1_{01}$). The spectrum was obtained at a pressure of 39 Pa (295 mTorr) in a 5 cm long cell. The residuals (x5) from the best fit MBE calculation are shown above (yellow) the magnitude spectrum (blue) b) The Fourier transformed RF comb spectrum (>3000 comb lines in 1$^{st}$ order) showing the carrier tone near 300 kHz and the (-) and (+) sideband comb spectra that span > 150 kHz intervals to the left and right of it, respectively. The inset shows the interleaving of the EOM orders on 10 Hz intervals and transform limited comb linewidth of $\approx$ 1 Hz. The signal-to-noise ratio of the magnitude spectrum is $\approx$ 20:1.

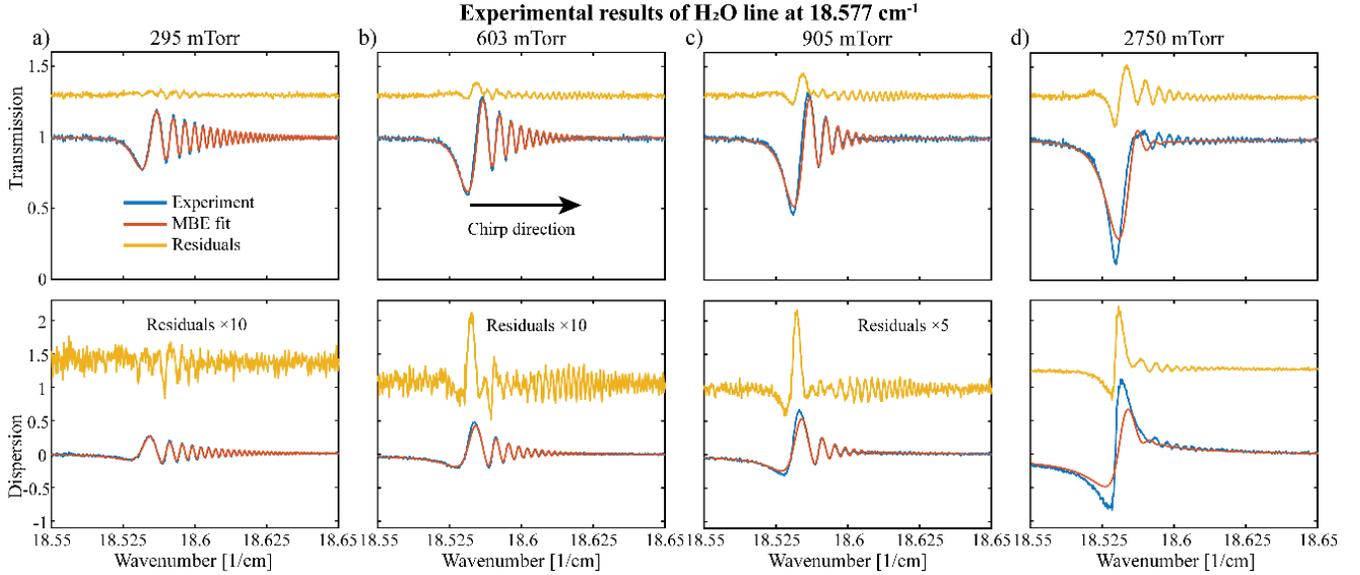

**Figure 3**. The transmission and dispersion response of $H_2O$ line at 557 GHz ($1_{01} \leftarrow 1_{10}$) at a) 39 Pa (295 mTorr), b) 80 Pa (603 mTorr), c) 121 Pa (905 mTorr), and d) 367 Pa (2750 mTorr) are shown in blue. The MBE fits are in red with residuals shown above in yellow.

The DF-EO-DOFC method is used to measure two pure rotational lines of $H_2O$ and $D_2O$. The ECDL laser was fixed near 11860 cm$^{-1}$ for $H_2O$ and 11864 cm$^{-1}$ for $D_2O$ with the Ti:Sapp frequency fixed near 11878 cm$^{-1}$. Figure 2a shows the transmission spectra (blue) of the $1_{1,0} \leftarrow 1_{0,1}$ transition of $H_2O$ at 18.577 cm$^{-1}$ obtained at 39 Pa (295 mTorr), a 100-record average. An example of the down-converted RF comb spectrum is shown in Fig. 2b for a 1 sec interferogram. The plus and minus sideband combs are displaced to either side of the $\Delta f_{AOM}$ = 300 kHz beat note. The inset shows the interleaving of the first two EOM orders on 10 Hz intervals and the transform limited 1 Hz width of the comb teeth. The separation between the comb teeth of a given order (not shown) is defined by the chirp repetition rate, $1/\tau_{CP}$ = 100 Hz.

In Fig. 2a, the elongated spectral response having post resonance oscillations is a clear signature of rapid passage effects. The frequency of the ripples depends on the chirp rate while the lifetime of the decaying envelope is defined by the dephasing time of the transition. As evident from the residuals (x5, yellow) shown above the magnitude spectrum in Fig. 2a, these effects are well modeled using the MBEs [23,24] that can describe the evolution of a two-level system [28] in the presence of a chirped field (details of the MBE method are given in the Supplement). To obtain satisfactory fits to these equations, the effective chirp rate used in the model is defined as,

$$\alpha_{MBE} = \left(\frac{\gamma}{2\pi} + 1\right)\alpha_{CR} \qquad (2)$$

where $\alpha_{CR}$ is the optical scan rate. Although the reasons for the additional factors of the $2\pi$ divisor and unity offset are not yet understood, identical factors were required to fit the near-IR data of $CO_2$ [10]. Nevertheless, the effective chirp rate of $\alpha_{MBE}$ = 10.01 MHz/ns quantitatively accounts for the rapid passage response shown in Fig 2a.

As discussed in previous work [29,30,31,32,33], the rapid passage response is pressure dependent. As the pressure is increased, the collisional dephasing time decreases which further dampens the envelope response function. The trend is illustrated in Fig. 3 which shows the observed transmission and dispersion spectra (100 record averages) of the 18.577 cm$^{-1}$ line at the four different pressures of 39 Pa, 80 Pa, 121 Pa and 367 Pa. At lower pressures shown in Figs. 3a and 3b, the extended ripples are well fit to the temporally magnified MBE model shown in red. However, as the pressure is increased in Figs. 3c and 3d, the residuals from the MBE fits are seen to significantly increase, indicating the model does not fully capture the dynamics of the molecular response.

To further explore the quantum dynamics with increasing pressure, water vapor is slowly leaked into the gas cell as data is continuously acquired to capture the real time evolution of the magnified rapid passage line shapes [34]. Data were acquired at a leak rate of approximately 0.5 Pa/s (4 mTorr/s) to cover a pressure range from 0.3 Pa to 600 Pa (2 mTorr to 4.5 Torr) in a 20 min period. Figure 4a shows the 2D results of a $H_2O$ pressure scan where each scan in the series represents a 5-record average. As apparent from the region marked with an arrow in Fig. 4a, the 180-degree phase flip moves closer to the initial resonance as the pressure is increased. The phase discontinuity is not captured in the magnified MBE model predictions shown in Fig. 4b and leads to the large residuals shown in Figs. 3c and 3d.

As a further test of the anomalous behavior observed for this particular $H_2O$ transition, similar 2D scans were performed for the $D_2O$ transition at 13.46 cm$^{-1}$. Pure $D_2O$ was leaked into the cell at approximately the same rate over a 20 min period and the results are shown in Fig. 4c for the same 5 record average. We note that due to the decreased line strength of the $D_2O$ line, the fractional transmission for $D_2O$ is near 85 % in contrast to only 10 % transmission for the $H_2O$ line. As for $H_2O$, the rapid passage responses of $D_2O$ transform nearly to the limit of the complex line shape functions as the pressure is increased to

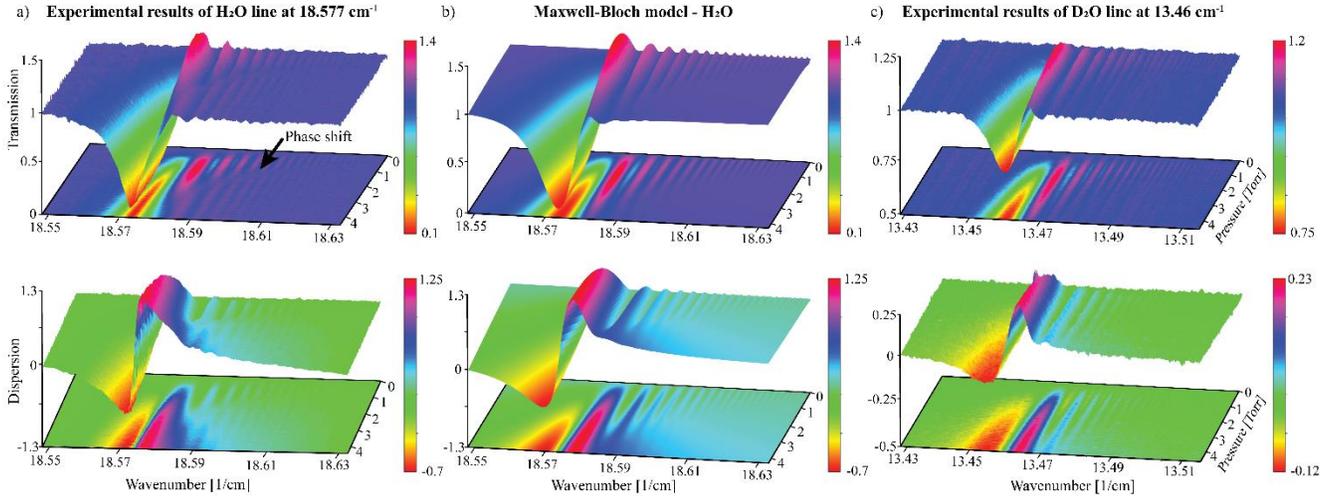

**Figure 4**. The evolution of the complex line shapes as the pressure is increased. Transmission (magnitude) and dispersion spectra are shown as surface plots with projections below. a) The experimental data for the $H_2O$ line at 18.577 cm$^{-1}$ as the pressure is increased from 0.2 Pa to 600 Pa (2 mTorr - 4.5 Torr). b) The simulated evolution of the magnified MBE line shape model of the $H_2O$ line at 18.577 cm$^{-1}$ corresponding to the data shown in (a). c) The experimental data of the $D_2O$ line at 13.46 cm$^{-1}$ as the pressure was increased to ≈ 600 Pa (4.5 Torr).

600 Pa (4.5 Torr). However, there is no indication of a phase flip or any anomalous behavior over the pressure range, marking a clear distinction between the $H_2O$ and $D_2O$ results.

The smooth pressure dependence of the phase flip in $H_2O$ argues for its origin in the spectroscopic properties of the transition. The $H_2O$ and $D_2O$ rotational transitions consist of a series of hyperfine components from the spin-rotation and spin-spin coupling between the rotational angular momentum, $J$, and the nuclear spins of the protons ($I$=1/2) or deuterons ($I$=1). The hyperfine transitions in $H_2O$ [35] and $D_2O$ [36] have been measured using a Lamb dip technique. For the $H_2O$ line, the principal hyperfine components consist of two transitions separated by ≈ 40 kHz while those of $D_2O$ consists mainly of 4 lines spread over a 100 kHz region. Because of the magnified relaxation time scale, a scan delay exists between the two hyperfine component absorptions in $H_2O$. Preliminary MBE simulations of these closely spaced resonances are shown to interfere over the expanded time scale which is consistent with the phase perturbation observed. However, unanswered questions that cast some doubt on this explanation include i) why the location of the phase flip following absorption depends on pressure and ii) why multiple phase perturbations are not observed in $D_2O$. Further experiments and modifications to the theory are warranted to arrive at a more satisfactory description of the anomalous quantum dynamics observed for this transition of $H_2O$.

In this work, we have demonstrated an interleaved chirped-pulse dual-optical-frequency-comb system that takes advantage of the mutual phase coherence between frequencies generated in two free-running lasers. While the laser linewidths are at the MHz level, the down-converted RF comb teeth display Hz (and even sub-Hz) linewidths in the RF region. The DF-EO-DOFC system has been used to investigate the complex transmission and dispersion profiles of pure rotational lines of $H_2O$ and $D_2O$ below 600 GHz. The dual chirped pulsed scheme is shown to transform these line shapes through temporal magnification to pure rapid passage signal responses. The $1_{10} \leftarrow 1_{01}$ transition of $H_2O$ exhibits a pressure-dependent phase flip in its rapid passage response that may depend on the interference between the two dominant hyperfine components. This anomalous behavior is, however, absent in the $2_{1,1} \leftarrow 2_{0,2}$ transition of $D_2O$, suggesting another mechanism may be responsible. The molecular dynamics demonstrated here using fixed-frequency continuous wave lasers were made possible only by the extreme temporal magnification of 62,893 generated using dual chirped-pulse waveforms. Current efforts are directed towards the rovibrational transitions of other gas phase molecules and THz phonon transitions of condensed phase biomolecular samples [21,37].

**Experimental methods:**
The lasers are free-space coupled into polarization maintaining fibers to give 120 mW in each fiber. Each of the two laser outputs is split into two equal legs. The ECDL outputs are then fed to two fiber coupled AOMs, Brimrose, TEM-50-2-60-850-2FP) operating near 50 MHz. The two AOMs, each with 3 dB insertion loss, are driven by two function generators (Keithley, 3390) at slightly different frequencies that are less than the 1 MHz RF bandwidth. The Ti:Sapp outputs are coupled to two EOMs (EO-space, PM-5SE-10-PFA-PFA-850-LV, 3 dB insertion loss), each driven by a channel of an arbitrary waveform generator (AWG, Keysight, M8195) that operates at a conversion rate of 32 GS/s each. Each EOM output is combined (50% / 50%) with one of the AOM outputs to define the Tx and Rx legs. The combined outputs are fiber coupled to a photomixer transmitter [13,17,18,19] (Tx, Toptica, EK-000685) and a photomixer receiver (Rx, Toptica, EK-000685). The laser powers in the Rx and Tx legs are approximately balanced and sum to 10-20 mW each.

The output of Tx is aligned directly to the Rx without a lens after passing through a 5 cm evacuable gas cell. The Tx is DC biased at 11 V and the electrical output current from the Tx is first amplified to voltage (Femto, DHPCA-100) and then digitized on an 8-bit digital oscilloscope (LeCroy, 813Zi). The amplifier bandwidth is gain dependent and typically operated at $10^6$ V/A to give a 3 dB roll-off at 1.8 MHz. A PCIe scope interface enables ≈ 80% throughput to the computer at 1 MS/sec. The two EOMs on the Tx and Rx legs are driven with chirped pulse waveforms defined according to the following [10],

$$WF_i^{Tx}(t) = \sin\left(2\pi f_{Tx_{start}} t + \frac{2\pi(f_{Tx_{stop}} - f_{Tx_{start}})}{2\tau_{CP}} t^2\right) \quad (3)$$

$$WF_i^{Rx}(t) = \sin\left(2\pi f_{Rx_{start}} t + \frac{2\pi(f_{Rx_{stop}} - f_{Rx_{start}})}{2\tau_{CP}} t^2 - \frac{2\pi}{N_{chirps}}\left(\frac{t}{\tau_{CP}} + i\right)\right) \quad (4)$$

$$i = 1 \dots N_{chirps}$$
$$t = 0 \dots \tau_{CP}$$

where $\tau_{CP}$ is the chirp duration and $N_{chirps}$ defines a frequency shift in the RF region that depends on the EOM order. When the quadratic term in Eq. 3 is non-zero, the optical and RF resolutions differ according to,

$$\Delta f_{Opt_{res}} = \left[1 - \frac{(f_{Rx_{stop}} - f_{Rx_{start}})}{(f_{Tx_{stop}} - f_{Tx_{start}})}\right]^{-1} \Delta f_{IF_{res}} \quad (5)$$

where $\Delta f_{IF_{res}}$ is 100 Hz (equal to $1/\tau_{CP}$).

The beat note, $\Delta f_{AOM} = 300$ kHz, is generated at the difference frequency of the AOMs to provide an offset to separate the EOM's plus and minus sidebands in the RF region while also moving the detection bandwidth away from low frequency noise. Additionally, the phase slip term, $\frac{2\pi}{N_{chirps}}\left(\frac{t}{\tau_{CP}} + i\right)$, in Eq. 4 is applied to the $N_{chirps}$ waveforms to define the fractional frequency shift that determines the interleaved locations of the different EOM orders in the RF spectrum [10]. The interleaving eliminates cross contamination between the EOM orders over a wide range of MW driving powers.

The waveforms used here are defined by $f_{Tx_{start}} = 0.5$ GHz and $f_{Tx_{stop}} = 10.5$ GHz on the Tx and $f_{Rx_{start}} = f_{Tx_{start}} - f_{IF_{start}}$ and $f_{Rx_{stop}} = f_{Tx_{stop}} - f_{IF_{stop}}$ on the Rx photomixer, where $f_{IF_{start}} = 5$ kHz and $f_{IF_{stop}} = 164$ kHz. In this case, the RF bandwidth (1.8 MHz) is nearly four orders-of-magnitude smaller than the AWG chirp bandwidth (10 GHz). Consequently, the magnification factor from Eq. 1 is 10 GHz/159 kHz = 62,893. With a chirp time of $\tau_{CP} = 10$ ms, the optical scan rate of the system is $\alpha_{CR} = 1\times10^{-3}$ MHz/ns. This scan rate is three decades lower than the usual threshold of ≈ 1 MHz/ns for observation of rapid passage effects [29,30,31,32,33]. The AWG waveform length is 100 ms to include the $N_{chirps} = 10$ phase shifted waveforms that interleave the EOM orders into 10 consecutive slots between RF comb teeth. To increase the transform limited resolution to 1 Hz, this waveform is repeated 10 times to generate a 1 sec-long interferogram record. Following the Fourier transformation, the RF comb lines, $f_{comb}(n, k)$, for the $k$-th EOM order are sampled to extract the transmission and dispersion spectra of the sample gas according to,

$$f_{comb}(n, k) = \pm \Delta f_{AOM} + n\Delta f_{IF_{res}} + \frac{k\Delta f_{IF_{res}}}{N_{chirps}} \quad (6)$$

The (±) comb lines for k=1 used in this work exist between $f_{IF_{start}}$ and $f_{IF_{stop}}$ (5 kHz and 164 kHz), on opposite sides of the beat note. Background spectra are acquired for an evacuated cell. Unlike the near-IR work, active phase stabilization of the free-running lasers was not used over the short 1 sec time records acquired here [5]. However, the random phase slip of the beat note between records required each interferogram to be properly paired with a compatible phase profile prior to normalization. A phase locking scheme is under development to coherently average over longer time periods.

**Funding**
Support was provided by the NIST Greenhouse Gas Measurements and Climate Research Program managed by Dr. James Whetstone.

**Acknowledgements**
J.R.S. wishes to acknowledge support from the NIST NRC fellowship program. We also wish to acknowledge helpful discussions with Jeeseong Hwang, Stephen McCoy, and Kimberly Briggman.

**Supporting Information Available**
A mathematical description of the Maxwell-Bloch equations used to model the rapid passage responses

**Data Availability**
Data for the results presented in this work are available at the following link.